# Large Magnetic-Field-Induced Strain at the Spin-Reorientation Transition in the A-Site Ordered Spinel Oxide LiFeCr$_4$O$_8$

Yoshihiko Okamoto[1,*], Tomoya Kanematsu[1], Yuki Kubota[1], Takeshi Yajima[2], and Koshi Takenaka[1]

[1]*Department of Applied Physics, Nagoya University, Nagoya 464-8603, Japan*
[2]*Institute for Solid State Physics, University of Tokyo, Kashiwa 277-8581, Japan*

Sintered samples of a spinel oxide LiFeCr$_4$O$_8$, where Cr$^{3+}$ and Fe$^{3+}$ ions have localized moments, were found to show a large magnetic-field-induced volume increase approaching 500 ppm by applying a magnetic field of 9 T. This large volume increase appeared only at ~30 K. At 30 K, a spin-reorientation transition from ferrimagnetic to conical order occurs, giving rise to this large volume increase. The coexistence of ferrimagnetic and conical phases at this transition was found to be important, suggesting that such a large magnetic-field-induced volume change can be realized at various magnetic transitions in localized magnets with strong spin–lattice coupling.

Chromium spinel compounds are known to be localized magnets that show various spin-lattice coupling phenomena.[1] They are insulator, in which Cr$^{3+}$ ions have $S = 3/2$ Heisenberg spins and form a pyrochlore structure consisting of corner-shared regular tetrahedra. In the case of oxides, where the antiferromagnetic interaction is dominant, the antiferromagnetic order is strongly suppressed by geometrical frustration of the pyrochlore structure. However, the frustration is always finally released by strong spin–lattice coupling, which results in long-range magnetic order accompanied by structural distortion.[2-4] This can be viewed as a spin analogue of the Jahn-Teller effect. In magnetic fields, Cr spinel compounds show a half magnetization plateau in their magnetization curves over a wide magnetic field range.[5,6] This is also accompanied by structural distortion caused by the strong spin-lattice coupling.[7,8]

Among the various spin-lattice coupling phenomena, Cr spinel compounds exhibit correlated phenomena between their magnetism and volume. This is unusual for a localized magnet, because such phenomena are mostly realized in itinerant magnets such as invar alloys.[9-12] For example, ZnCr$_2$Se$_4$ and LiGaCr$_4$S$_8$ showed negative thermal expansion above the Néel temperature, and LiInCr$_4$S$_8$ exhibited a substantial volume increase induced by applying a magnetic field.[13,14] In these materials, the competition between antiferromagnetic and ferromagnetic interactions as well as the strong spin–lattice coupling might be crucial.[15,16]

In this study, we focused on the *A*-site ordered Cr spinel oxide LiFeCr$_4$O$_8$, where Fe atoms have localized spins in addition to the Cr atoms. In LiFeCr$_4$O$_8$, nonmagnetic Li$^+$ and $S = 5/2$ Fe$^{3+}$ ions alternately occupy the tetrahedral sites (*A* sites), resulting in a zinc-blende-type configuration, as shown in the inset of Fig. 1(b). Cr$^{3+}$ ions with $S = 3/2$ spins occupy the octahedral sites and form a breathing pyrochlore structure, consisting of alternating small and large regular tetrahedra.[17] LiInCr$_4$O$_8$ and LiGaCr$_4$O$_8$, where nonmagnetic In$^{3+}$ and Ga$^{3+}$ ions occupy tetrahedral sites instead of Fe$^{3+}$ ions, respectively, are known to exhibit various spin-lattice coupling phenomena, as in the case of ordinary Cr spinel oxides.[18-23] LiFeCr$_4$O$_8$ was found to be insulating and to have a Weiss temperature $\theta_W$ greater than −1000 K, implying the presence of a very strong antiferromagnetic interaction.[24] In LiFeCr$_4$O$_8$, in addition to the antiferromagnetic interaction between Cr$^{3+}$ spins, magnetic interactions between Cr$^{3+}$ and Fe$^{3+}$ spins are strongly antiferromagnetic, probably resulting in the very large $|\theta_W|$. At $T_N = 94$ K, LiFeCr$_4$O$_8$ was reported to show a collinear ferrimagnetic order, with Cr$^{3+}$ and Fe$^{3+}$ spins aligned antiparallel. A spin-reorientation transition from collinear ferrimagnetic order to conical spin structure occurred at $T_{MS} = 23$ K. The transition at $T_{MS}$ was accompanied by a symmetry lowering from cubic $F$–43$m$ to tetragonal $I$–4$m$2. No dilatometric measurements have been made for LiFeCr$_4$O$_8$.

Here, we report that sintered samples of LiFeCr$_4$O$_8$ showed large volume increase of up to $\Delta V/V = 490$ ppm only at around $T_{MS}$ by applying a magnetic field of 9 T. This volume change was not completely restored when the magnetic field was reduced to zero, but it was initialized when the sample was cooled to well below $T_{MS}$, implying that the two-phase coexisting state of the ferrimagnetic and conical phases at the spin-reorientation transition was crucial in the large volume change. This behavior is a spin-reorientation variation of the large volume change in AgCrS$_2$. AgCrS$_2$ behaved similarly at the Néel temperature.[25] These findings suggested that a large magnetic-field-induced volume change can occur at various magnetic transitions in Cr-based localized magnets, which might be promising as a giant magnetostrictive material.

Sintered samples of LiFeCr$_4$O$_8$ were prepared by solid-state reaction method.[26] Linear thermal expansion and linear strain in magnetic fields of the sintered samples were measured using a strain gage (KFLB, 120 W, Kyowa Electronic Instruments Co.) with a Cu reference.[27] In the isotropic situation, the volume change $\Delta V/V$ and linear strain $\Delta L/L$ have a relationship of $\Delta V/V = 3(\Delta L/L)$ in the linear thermal expansion measurements. In an isotropic material, however, a volume change in a magnetic field is represented as $\Delta V/V = 2(\Delta L/L)_\perp + (\Delta L/L)_{//}$, where $(\Delta L/L)_\perp$ and $(\Delta L/L)_{//}$ are the linear strains measured perpendicular and parallel to the magnetic fields, respectively. The linear strain under a magnetic field and the thermal expansion were measured using a Physical Property Measurement System (Quantum Design). Powder X-ray diffraction (XRD) patterns of LiFeCr$_4$O$_8$ samples at various temperatures were measured using a Miniflex diffractometer with Cu Kα radiation and a SmartLab diffractometer (both RIGAKU) with Cu Kα$_1$ radiation, the latter of which was monochromated by a Ge(111)-Johansson-type monochromator. Magnetization was measured using a Magnetic Property Measurement System (Quantum Design). Heat capacity was measured by a relaxation method using a Physical Property Measurement System. The crystal structure image was created using VESTA.[28]

The magnetization $M$ of the LiFeCr$_4$O$_8$ sintered sample sharply increased with decreasing temperature below 100 K, accompanied by temperature hysteresis, as shown in Fig. 1(a). At lower temperatures, $M$ showed a maximum at 50 K for the field-cooled data and at 60 K for the zero-field-cooled data, followed by a strong decrease above 30 K. Both data showed a kink at 30 K and were almost constant below this temperature. This temperature dependence was similar to that reported in a previous study,[24] showing that the ferrimagnetic order and the spin-reorientation transition from ferrimagnetic to conical order occurred at $T_N$ = 100 K and $T_{MS}$ = 30 K, respectively. The temperature dependence of the heat capacity divided by temperature, $C_p/T$, of a LiFeCr$_4$O$_8$ sintered sample, shown in Fig. 1(b), showed a large peak at $T_{MS}$ = 30 K. This supported the presence of the spin-reorientation transition at this temperature. The value of $T_{MS}$ = 30 K was also consistent with the thermal expansion data presented below. As seen in Fig. 1(b), there is a minor peak corresponding to $T_N$ in the $C_p/T$ data. These $T_{MS}$ and $T_N$ values are several K higher than those reported in the previous study,[24] which might be due to the difference in the minute amount of intersite mixing and defects between these samples. There was a small anomaly at 60 K in the $C_p/T$ data of the previous study,[24] which also appeared in the $C_p/T$ data in Fig. 1(b), as indicated by $T^*$. As discussed below, this anomaly also appeared in the linear thermal expansion data. Unlike $T_{MS}$ and $T_N$, however, the $T^*$ values in this and previous studies are identical, implying that this anomaly might not be intrinsic to LiFeCr$_4$O$_8$.

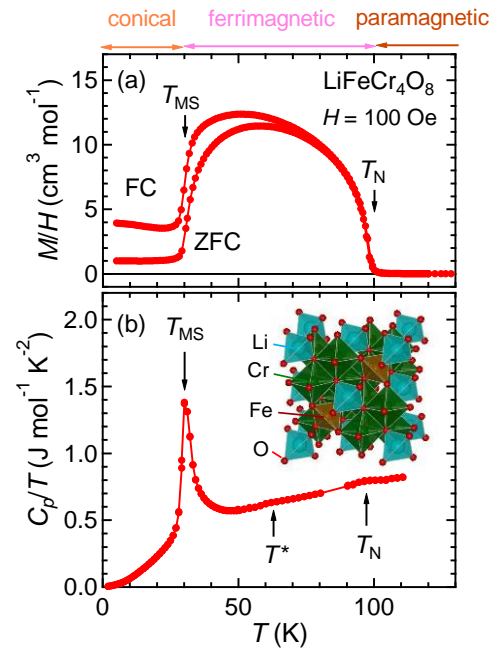

Figure 1. (a) Temperature dependence of field-cooled and zero-field-cooled magnetization divided by the magnetic field $M/H$ of the LiFeCr$_4$O$_8$ sintered sample measured at a magnetic field of 100 Oe. (b) Temperature dependence of heat capacity divided by temperature of the LiFeCr$_4$O$_8$ sintered sample. Magnetic structures shown above (a) were reported in a previous study.[24] The inset of (b) shows the crystal structure of LiFeCr$_4$O$_8$.

The linear thermal expansion, $\Delta L/L$, of the sintered sample of LiFeCr$_4$O$_8$ is shown in Fig. 2. For comparison, the data for LiInCr$_4$O$_8$ and LiGaCr$_4$O$_8$ are also shown.[14] The $\Delta L/L$ of LiFeCr$_4$O$_8$ was almost identical to that of LiInCr$_4$O$_8$ in the temperature range of ambient temperature and 250 K. At lower temperature, their $\Delta L/L$ data deviated from each other and the coefficient of thermal expansion, α, of LiFeCr$_4$O$_8$ became small in the ferrimagnetic phase between $T_N$ = 100 K and $T_{MS}$ = 30 K, with α = 1.5 × 10$^{-6}$ between 60 and 100 K. This α is quite small for transition metal oxides, implying that the volume increase associated with the development of ferrimagnetic order is superimposed over the normal thermal expansion. The volume change of $\Delta V/V$ = 2600 ppm between 100 and 300 K was consistent with that calculated using the lattice constants determined by powder XRD data of $\Delta V/V$ = 2100 ppm. There was also a small anomaly in the $\Delta L/L$ data at $T^*$ = 60 K, which most likely matched to the small anomaly at $T^*$ in the $C_p/T$ data shown in Fig. 1(b).

At $T_{MS}$ = 30 K, the $\Delta L/L$ of LiFeCr$_4$O$_8$ sharply decreased with decreasing temperature, which is due to the structural change at the spin reorientation transition from the collinear ferrimagnetic to conical. The volume change at $T_{MS}$ was $\Delta V/V = 3(\Delta L/L) = 1200$ ppm. The volume decrease associated with symmetry lowering of the crystal structure is common for Cr spinels.[14] The change of $\Delta L/L$ at $T_{MS}$ was accompanied by temperature hysteresis of ~2 K, indicating the first-order phase transition, as seen in the inset of Fig. 2. The $\Delta L/L$ data measured in a magnetic field of 5 T showed almost the same change at $T_{MS}$ as that of 0 T, but the transition temperature was 1 K lower.

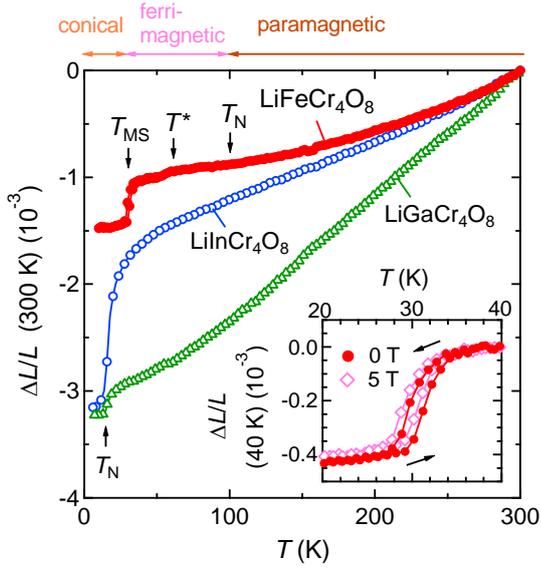

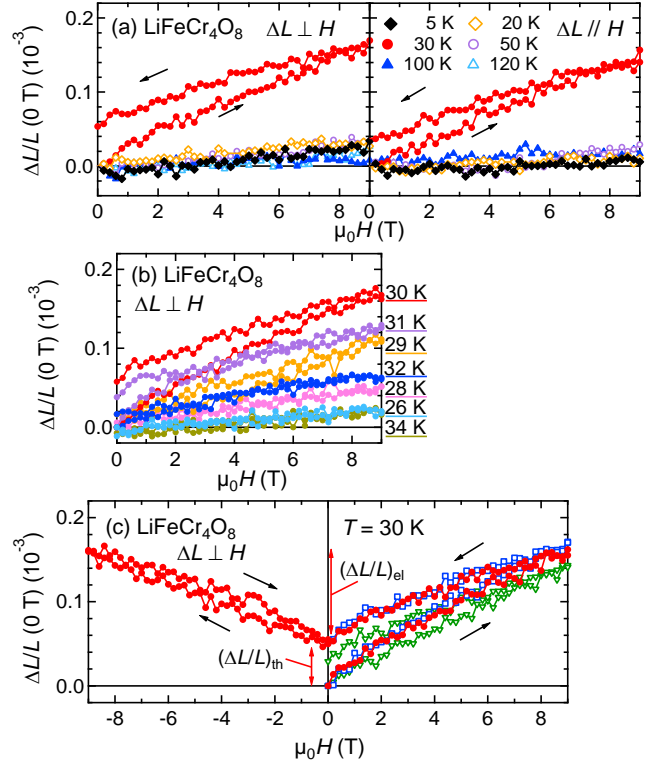

Figure 2. Linear thermal expansion of the LiFeCr$_4$O$_8$ sintered sample normalized to 300 K. Data for the LiInCr$_4$O$_8$ and LiGaCr$_4$O$_8$ sintered samples are also shown for comparison.[14] The inset shows the data around $T_N$ normalized to 40 K. In the inset, the data taken at 0 (filled) and 5 T (open) are shown. Magnetic structures were reported in a previous study.[24]

The magnetic field dependences of the linear strains of a sintered sample of LiFeCr$_4$O$_8$ are shown in Fig. 3. At 29, 30, and 31 K, large strains of $\Delta L/L > 100$ ppm were found in the magnetic field of 9 T, although strains at other temperatures were negligible, several tens of ppm at most. This contrasted with the absence of large magnetic-field-induced strain in LiInCr$_4$O$_8$ and LiGaCr$_4$O$_8$.[14] When the magnetic field was reduced to zero, the $\Delta L/L$ at ~30 K was not completely recovered, as will be discussed in more detail. The linear strains from 0 to 9 T at 30 K were $(\Delta L/L)_\perp = 170$ ppm and $(\Delta L/L)_{//} = 150$ ppm, which were almost isotropic, providing a large volume increase of $\Delta V/V = 490$ ppm. Although this $\Delta V/V$ was smaller than the maximum value of $\Delta V/V = 780$ ppm in the isostructural sulfide LiInCr$_4$S$_8$,[14] it was still quite large for a localized spin system. Furthermore, the almost isotropic strain demonstrated that the significant strain was not caused by the rearrangement of ferrimagnetic domains. This was supported by the fact that large strains did not appear at 50 and 34 K in the ferrimagnetic phase.

Figure 3. Linear strains of a sintered sample of LiFeCr$_4$O$_8$ measured in magnetic fields. All data were normalized to the strain data taken at 0 T. (a) Linear strains measured perpendicular to (left) and along (right) the magnetic fields. After keeping the sample at 5 K under zero magnetic field, the 5 K data were measured first, followed by the other data in order of increasing temperature. (b) Linear strains perpendicular to the magnetic fields between 26 and 34 K. After the sample was warmed from 5 K at 0 T, the linear strain was measured with increasing magnetic field from 0 to 9 T and decreasing magnetic field from 9 to 0 T at each temperature. (c) Linear strains perpendicular to the magnetic fields at 30 K. Filled circles show the linear strain measured after keeping the sample at 5 K in the zero magnetic field. Open triangles show the linear strain after the data indicated by filled circles were measured and then the sample was kept at 120 K. Open squares show the linear strain after the data indicated by open triangles were measured and then the sample was kept at 5 K.

Thus, the LiFeCr$_4$O$_8$ sintered sample showed substantial magnetic-field-induced volume changes only at around $T_{MS}$. Even when the magnetic field was removed, the volume changes were not restored. As shown in Fig. 3(c), however, the volume change was initialized by maintaining the sample at 5 K and then warming it to 30 K again (open squares), rather than by keeping the sample at high temperature (open triangles). This result revealed that the large volume change of LiFeCr$_4$O$_8$ comprised both elastic and thermoelastic contributions, which were appeared as $(\Delta L/L)_{el}$ and $(\Delta L/L)_{th}$ in Fig. 3(c), respectively. These characteristics distinguished LiFeCr$_4$O$_8$ from other Cr spinels, such as LiInCr$_4$S$_8$, ZnCr$_2$Se$_4$, and FeCr$_2$O$_4$, which showed large magnetic-field-induced volume changes or strains.[13,14,29] The large volume changes or strains in LiInCr$_4$S$_8$ and ZnCr$_2$Se$_4$ were elastic and appeared below the magnetic order temperature, reflecting that the magnetic-field-induced phase transition was responsible for them. Single crystals of FeCr$_2$O$_4$ with both Fe$^{2+}$ and Cr$^{3+}$ magnetic ions showed a very large linear strain of up to $\Delta L/L$ = 3000 ppm by applying a magnetic field of 7 T in the ferrimagnetically ordered phase.[29] The Fe$^{2+}$ ion in FeCr$_2$O$_4$ is Jahn-Teller active in contrast to Fe$^{3+}$ in LiFeCr$_4$O$_8$. Rearrangement of the ferrimagnetic domains, in combination with the Jahn-Teller distortion in magnetic fields, resulted in a large strain. The substantial volume changes in LiFeCr$_4$O$_8$ were distinct from those in these three Cr spinel compounds, implying that the formation mechanisms were different, but similar to those in a triangular antiferromagnet AgCrS$_2$,[25] as discussed below.

The presence of thermoelastic volume change, as well as the fact that the large volume changes appeared in the hysteresis loop of linear thermal expansion data in the inset of Fig. 2, strongly suggested that the coexistence of the ferrimagnetic and conical phases at $T_{MS}$ was crucial in the large magnetic-field-induced volume change. As discussed above, the volume of the ferrimagnetic phase above $T_{MS}$ was $\Delta V/V$ = 1200 ppm larger than the volume of the conical phase below $T_{MS}$. When the sample was warmed from 5 to 30 K, almost all the sample was in the conical phase. By applying a magnetic field, the sample partly changed to the ferrimagnetic phase with a larger volume, giving rise to a large volume increase. Because this was not a phase transition between thermal equilibrium states, unlike ZnCr$_2$Se$_4$ and LiInCr$_4$S$_8$, the volume change could be irreversible. Furthermore, when the magnetic field was reduced to zero, part of the sample became conical, while the rest remained ferrimagnetic. The former produces $(\Delta L/L)_{el}$, while the latter produces $(\Delta L/L)_{th}$. This $(\Delta L/L)_{th}$ was initialized by holding the sample at a low enough temperature to convert all of the sample to the conical phase.

These features of the magnetic-field-induced volume change of LiFeCr$_4$O$_8$ are similar to those of the triangular antiferromagnet AgCrS$_2$.[25] At $T_N$ = 42 K, AgCrS$_2$ showed an antiferromagnetic order with a large volume increase of $\Delta V/V$ = 1800 ppm in a sintered sample, indicating the presence of a strong spin–lattice coupling, similar to the Cr spinel compounds. The crystal structures of Cr spinel compounds and AgCrS$_2$ consist of edge-shared octahedra coordinating to Cr$^{3+}$ ions, which is expected to be the source of their strong spin–lattice coupling. The sintered samples of AgCrS$_2$ exhibited a large volume change in magnetic fields only at $T_N$ = 42 K. This large volume change was caused by the increase of a volume fraction of a paramagnetic phase with a larger volume in the two-phase coexisting state at $T_N$. Because AgCrS$_2$ is a sulfide, it has a strong ferromagnetic interaction between the Cr$^{3+}$ spins in addition to the antiferromagnetic interaction.[30,31] The presence of the strong ferromagnetic interaction has been suggested to play a role in the increase of the paramagnetic fraction in magnetic fields.[25] In contrast, the antiferromagnetic interaction is dominant in the oxides, such as LiFeCr$_4$O$_8$, LiInCr$_4$O$_8$, and LiGaCr$_4$O$_8$.[15,17,24] In fact, LiInCr$_4$O$_8$ and LiGaCr$_4$O$_8$ showed an antiferromagnetic order with a considerable volume change at $T_N$ = 15 K, probably reflecting the presence of strong spin–lattice coupling, although there is no significant magnetic-field-induced volume change at any temperature, including $T_N$.[14] In the two-phase coexisting state of the ferrimagnetic and conical phases in LiFeCr$_4$O$_8$, on the other hand, the ferrimagnetic phase with a much larger magnetization is expected to be more stable in magnetic fields than the conical phase. The increase in the fraction of the ferrimagnetic phase with a larger volume in magnetic fields probably appeared as a large volume increase.

The above discussion suggests that spin-reorientation transitions with volume changes are a novel mechanism for producing a large magnetic-field-induced volume change in a localized spin system. Giant magnetostrictive materials, such as Terfenol-D, which are itinerant ferromagnets and exhibit large strains due to the alignment of ferromagnetic domains in magnetic fields, have been discovered to date.[32] The results of this study did not only demonstrated a wide variety of magnetovolume phenomena in Cr spinel compounds, but also indicated that the substantial magnetic-field-induced volume change could appear at various magnetic transitions in localized magnets with strong spin–lattice coupling, which could considerably expand the target materials for the creation of next-generation magnetostrictive materials.

In summary, we found that a sintered sample of the $A$-site ordered spinel oxide LiFeCr$_4$O$_8$ exhibited a large magnetic-field-induced volume change approaching 500 ppm in a magnetic field of 9 T at $T_{MS}$ = 30 K, where the spin-reorientation transition from the high-temperature ferrimagnetic to the low-temperature conical phases occurred. This large volume change appeared only at $T_{MS}$ and was not restored after the magnetic field was removed, but it was initialized by keeping the sample to a temperature well below $T_{MS}$. The phase transition at $T_{MS}$ was first order with a temperature hysteresis and the volume of the ferrimagnetic phase was $\Delta V/V$ = 1200 ppm larger than the volume of the conical phase, according to the linear thermal expansion data. These findings strongly suggested that the large magnetic-field-induced volume change in LiFeCr$_4$O$_8$ was caused by the increase of the volume fraction of the ferrimagnetic phase in the two-phase coexisting state of ferrimagnetic and conical phases at $T_{MS}$.

## ACKNOWLEDGMENTS

The authors are grateful to T. Yamauchi for his support with the heat capacity measurements. The work was partly carried out under the Visiting Researcher Program of the Institute for Solid State Physics, the University of Tokyo and partly supported by the Collaborative Research Project of Materials and Structures Laboratory, Tokyo Institute of Technology and JSPS KAKENHI (Grant Numbers: 19H05823, 20H00346, and 20H02603).

1) S.-H. Lee, H. Takagi, D. Louca, M. Matsuda, S. Ji, H. Ueda, Y. Ueda, T. Katsufuji, J.-H. Chung, S. Park, S.-W. Cheong, and C. Broholm, J. Phys. Soc. Jpn. **79**, 011004 (2010).
2) H. Ueda, H. Mitamura, T. Goto, and Y. Ueda, Phys. Rev. B **73**, 094415 (2006).
3) N. Shannon, H. Ueda, Y. Motome, K. Penc, H. Shiba, and H. Takagi, J. Phys. Conf. Ser. **51**, 31 (2006).
4) S.-H. Lee, C. Broholm, T. H. Kim, W. Ratcliff II, and S.-W. Cheong, Phys. Rev. Lett. **84**, 3718 (2000).
5) H. Ueda, H. A. Katori, H. Mitamura, T. Goto, and H. Takagi, Phys. Rev. Lett. **94**, 047202 (2005).
6) H. Ueda, H. Mitamura, T. Goto, and Y. Ueda, Phys. Rev. B **73**, 094415 (2006).
7) M. Matsuda, H. Ueda, A. Kikkawa, Y. Tanaka, K. Katsumata, Y. Narumi, T. Inami, Y. Ueda, and S.-H. Lee, Nat. Phys. **3**, 397 (2007).
8) M. Matsuda, K. Ohoyama, S. Yoshii, H. Nojiri, P. Frings, F. Duc, B. Vignolle, G. L. J. A. Rikken, L.-P. Regnault, S.-H. Lee, H. Ueda, and Y. Ueda, Phys. Rev. Lett. **104**, 047201 (2010).
9) C. E. Guillaume, CR Acad. Sci. **125**, 235 (1897).
10) C. E. Guillaume, Compt. Rend. **171**, 1039 (1920).
11) H. Nagaoka and K. Honda, Phil. Mag. **4**, 45 (1902).
12) M. Matsumoto, T. Kaneko, and H. Fujimori, J. Phys. Soc. Jpn. **26**, 1083 (1969).
13) J. Hemberger, H.-A. Krug von Nidda, V. Tsurkan, and A. Loidl, Phys. Rev. Lett. **98**, 147203 (2007).
14) T. Kanematsu, M. Mori, Y. Okamoto, T. Yajima, and K. Takenaka, J. Phys. Soc. Jpn. **89**, 073708 (2020).
15) P. Ghosh, Y. Iqbal, T. Müller, R. T. Ponnaganti, R. Thomale, R. Narayanan, J. Reuther, M. J. P. Gingras, and H. O. Jeschke, npj Quantum Mater. **4**, 63 (2019).
16) Y. Okamoto, M. Mori, N. Katayama, A. Miyake, M. Tokunaga, A. Matsuo, K. Kindo, and K. Takenaka, J. Phys. Soc. Jpn. **87**, 034709 (2018).
17) Y. Okamoto, G. J. Nilsen, J. P. Attfield, and Z. Hiroi, Phys. Rev. Lett. **110**, 097203 (2013).
18) G. J. Nilsen, Y. Okamoto, T. Matsuda, J. Rodriguez-Carvajal, H. Mutka, T. Hansen, and Z. Hiroi, Phys. Rev. B **91**, 174435 (2015).
19) S. Lee, S.-H. Do, W.-J. Lee, Y. S. Choi, M. Lee, E. S. Choi, A. P. Reyes, P. L. Kuhns, A. Ozarowski, and K.-Y. Choi, Phys. Rev. B **93**, 174402 (2016).
20) R. Saha, F. Fauth, M. Avdeev, P. Kayser, B. J. Kennedy, and A. Sundaresan, Phys. Rev. B **94**, 064420 (2016).
21) R. Wawrzyńczak, Y. Tanaka, M. Yoshida, Y. Okamoto, P. Manuel, N. Casati, Z. Hiroi, M. Takigawa, and G. J. Nilsen, Phys. Rev. Lett. **119**, 087201 (2017).
22) Y. Okamoto, D. Nakamura, A. Miyake, S. Takeyama, M. Tokunaga, A. Matsuo, K. Kindo, and Z. Hiroi, Phys. Rev. B **95**, 134438 (2017).
23) M. Gen, D. Nakamura, Y. Okamoto, and S. Takeyama, J. Magn. Magn. Mater. **473**, 387 (2019).
24) R. Saha, R. Dhanya, C. Bellin, K. Béneut, A. Bhattacharyya, A. Shukla, C. Narayana, E. Suard, J. Rodríguez-Carvajal, and A. Sundaresan, Phys. Rev. B **96**, 214439 (2017).
25) T. Kanematsu, Y. Okamoto, and K. Takenaka, Appl. Phys. Lett. **118**, 142404 (2021).
26) The details on the sample preparation are provided in Supplementary Note.
27) K. Takenaka, M. Ichigo, T. Hamada, A. Ozawa, T. Shibayama, T. Inagaki, and, K. Asano, Sci. Technol. Adv. Mater. **15**, 015009 (2014).
28) K. Momma and F. Izumi, J. Appl. Crystallogr. **44**, 1272 (2011).
29) H. Sagayama, S. Ohtani, M. Saito, N. Abe, K. Taniguchi, and T. Arima, Appl. Phys. Lett. **99**, 082506 (2011).
30) A. V. Ushakov, D. A. Kukusta, A. N. Yaresko, and D. I. Khomskii, Phys. Rev. B **87**, 014418 (2013).
31) F. Damay, C. Martin, V. Hardy, G. André, S. Petit, and A. Maignan, Phys. Rev. B **83**, 184413 (2011).
32) A. E. Clark, *Handbook of Ferromagnetic Materials* (North-Holland Publishing Company, Amsterdam, 1980), Vol. 1, pp. 531-589.

*E-mail: yokamoto@nuap.nagoya-u.ac.jp

**Supplementary Note 1. Sample preparation and powder X-ray diffraction data of LiFeCr$_4$O$_8$.**

Sintered samples of LiFeCr$_4$O$_8$ were prepared by the solid-state reaction method. Li$_2$CO$_3$, Fe$_2$O$_3$, and Cr$_2$O$_3$ powders were mixed in a 1.03:1:4 molar ratio. The mixture was pressed into pellets, put in an alumina crucible, and then heated to and kept at 1273 K for 30 h in air with an intermediate grinding. Supplementary Fig. 1 shows a powder X-ray diffraction pattern of a sintered sample of LiFeCr$_4$O$_8$ measured at room temperature. All the peaks, except for the small peaks from the Cr$_2$O$_3$ impurity phase, were indexed on the basis of a cubic unit cell with the space group of *F*–43*m* and a lattice constant of *a* = 8.2753(3) Å. This *a* was almost the same as that reported in a previous study [1], indicating that the obtained sample is almost entirely single-phase LiFeCr$_4$O$_8$.

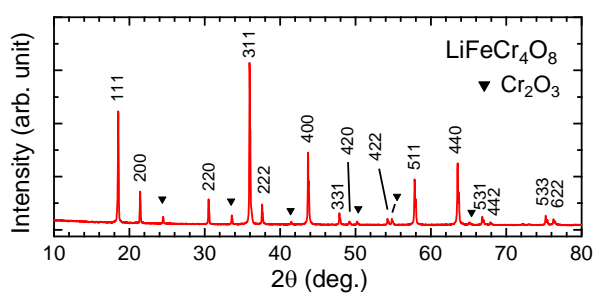

Supplementary Figure 1. Powder XRD pattern of the LiFeCr$_4$O$_8$ polycrystalline sample measured at room temperature using Cu Kα radiation. Peaks indicated by triangles are those of a Cr$_2$O$_3$ impurity. Peak indices are given using a cubic unit cell with a lattice constant of *a* = 8.2753(3) Å.

[1] R. Saha, R. Dhanya, C. Bellin, K. Béneut, A. Bhattacharyya, A. Shukla, C. Narayana, E. Suard, J. Rodríguez-Carvajal, and A. Sundaresan, Phys. Rev. B **96**, 214439 (2017).